\documentclass[cits]{PoS}

\usepackage{amsmath}

\title{Polyakov loops and SU(2) staggered Dirac spectra}

\ShortTitle{Polyakov loops and SU(2) staggered Dirac spectra}

\author{Falk Bruckmann$^a$, Stefan Keppeler$^{a,b}$, \speaker{Marco
    Panero}$^a$, and Tilo Wettig$^a$\\ 
  $^a$Institut f\"ur Theoretische Physik, Universit\"at Regensburg,
  93040 Regensburg, Germany\\
  \mbox{$^b$Mathematisches Institut, Universit\"at T\"ubingen,
    Auf der Morgenstelle 10, 72076 T\"ubingen, Germany}\\
  E-mail: \email{marco.panero@physik.uni-regensburg.de}}

\abstract{We consider the spectrum of the staggered Dirac operator
  with SU(2) gauge fields.  Our study is motivated by the fact that
  the antiunitary symmetries of this operator are different from those
  of the SU(2) continuum Dirac operator.  In this contribution, we
  investigate in some detail staggered eigenvalue spectra close to the
  free limit.  Numerical experiments in the quenched approximation and
  at very large $\beta$-values show that the eigenvalues occur in
  clusters consisting of eight eigenvalues each.  We can predict the
  locations of these clusters for a given configuration very
  accurately by an analytical formula involving Polyakov loops and
  boundary conditions.  The spacing distribution of the eigenvalues
  within the clusters agrees with the chiral symplectic ensemble of
  random matrix theory, in agreement with theoretical expectations,
  whereas the spacing distribution between the clusters tends towards
  Poisson behavior.}

\FullConference{The XXV International Symposium on Lattice Field Theory\\
  July 30 - August 4, 2007\\
  Regensburg, Germany}

\begin{document}

\section{Introduction}

Random matrix theory (RMT)~\cite{RMT} accurately describes eigenvalue
correlations in complex systems.  More precisely, if we consider a
quantum system governed by a Hamiltonian $H$ whose classical
counterpart is chaotic, then the statistical properties of the
eigenvalue spectrum of $H$ can be modeled by an ensemble of matrices
with random entries (distributed according to some statistical weight)
and with the same global symmetries as $H$. This description is
insensitive to the details of the interaction and predicts universal
features that are unveiled when different spectra are rescaled
(unfolded) to the same mean density.

Among the many applications of RMT in mathematics and in physics, a
particularly interesting one is relevant for the description of the
spectrum of the Dirac operator in quantum chromodynamics.  QCD
in the $\varepsilon$-regime can be described by a chiral RMT with the
same chiral and flavor symmetries as QCD~\cite{chRMT}.  This approach
can also be extended to non-vanishing temperature and/or chemical
potential and has been confirmed in many numerical studies.  In this
formulation, the anti-Hermitian massless Dirac operator $D=\gamma_\mu
\left(\partial_\mu + i A_\mu \right)$ is described in terms of a
matrix with an off-diagonal block structure,
\begin{equation}
  \label{dmatrix}
  D\to \left( 
    \begin{tabular}{cc}
      0 & $i W$ \\
      $i W^\dagger$ & 0 \\
    \end{tabular}
  \right) \: , 
\end{equation}
where $W$ is a complex $(n+\nu)\times n$ matrix and $\nu$ plays the
role of the topological charge.

Depending on the color gauge group $G$ and on the fermion field
representation, the Dirac operator may also be invariant under some
discrete antiunitary symmetries, leading to the following symmetry
classes~\cite{Verbaarschot:1994qf}:
\begin{enumerate}
\item For $G=\text{SU}(2)$ and fermions in the fundamental
  representation, the pseudo-real nature of the group generators
  allows us to recast the Dirac operator in a form with real matrix
  entries. The corresponding matrix ensemble is the chiral orthogonal
  ensemble (chOE) with Dyson index $\beta_D=1$.
\item For $G=\text{SU}(N_C)$ with $N_C \ge 3$ and fermions in the
  fundamental representation, the Dirac operator generically has
  complex entries.  The appropriate matrix ensemble is the chiral
  unitary ensemble (chUE) with Dyson index $\beta_D=2$.
\item For gauge group $G=\text{SU}(N_C)$ and fermions in the adjoint
  representation, the generators are antisymmetric matrices with
  imaginary entries, and the Dirac operator can be written as a matrix
  of real quaternions. The associated matrix ensemble is the chiral
  symplectic ensemble (chSE) with Dyson index $\beta_D=4$.
\end{enumerate}
The behavior of the universal quantities depends on the symmetry
classes listed above.  In particular, the probability density $P(s)$
for the spacing $s$ of adjacent unfolded levels can be computed
exactly and is well approximated by the Wigner surmise,
\begin{align}
  \label{eq:Wigner}
  P(s) = a\,s^{\beta_D}e^{-bs^2}\quad\text{with}\quad
  a=2\,\frac{\Gamma^{\beta_D+1} \left( \beta_D/2 +1 \right)}
  {\Gamma^{\beta_D+2} \left( (\beta_D+1)/2 \right)} \:,\quad
  b= \frac{\Gamma^2\left( \beta_D/2+1 \right)}
    {\Gamma^2 \left( (\beta_D +1)/2 \right)}\:.
\end{align}
For quantum systems whose classical analog is integrable, $P(s)$ is
given by the result for a Poisson process, $P(s)=e^{-s}$.

\begin{figure}[-t]
  \includegraphics[width=.33\textwidth]{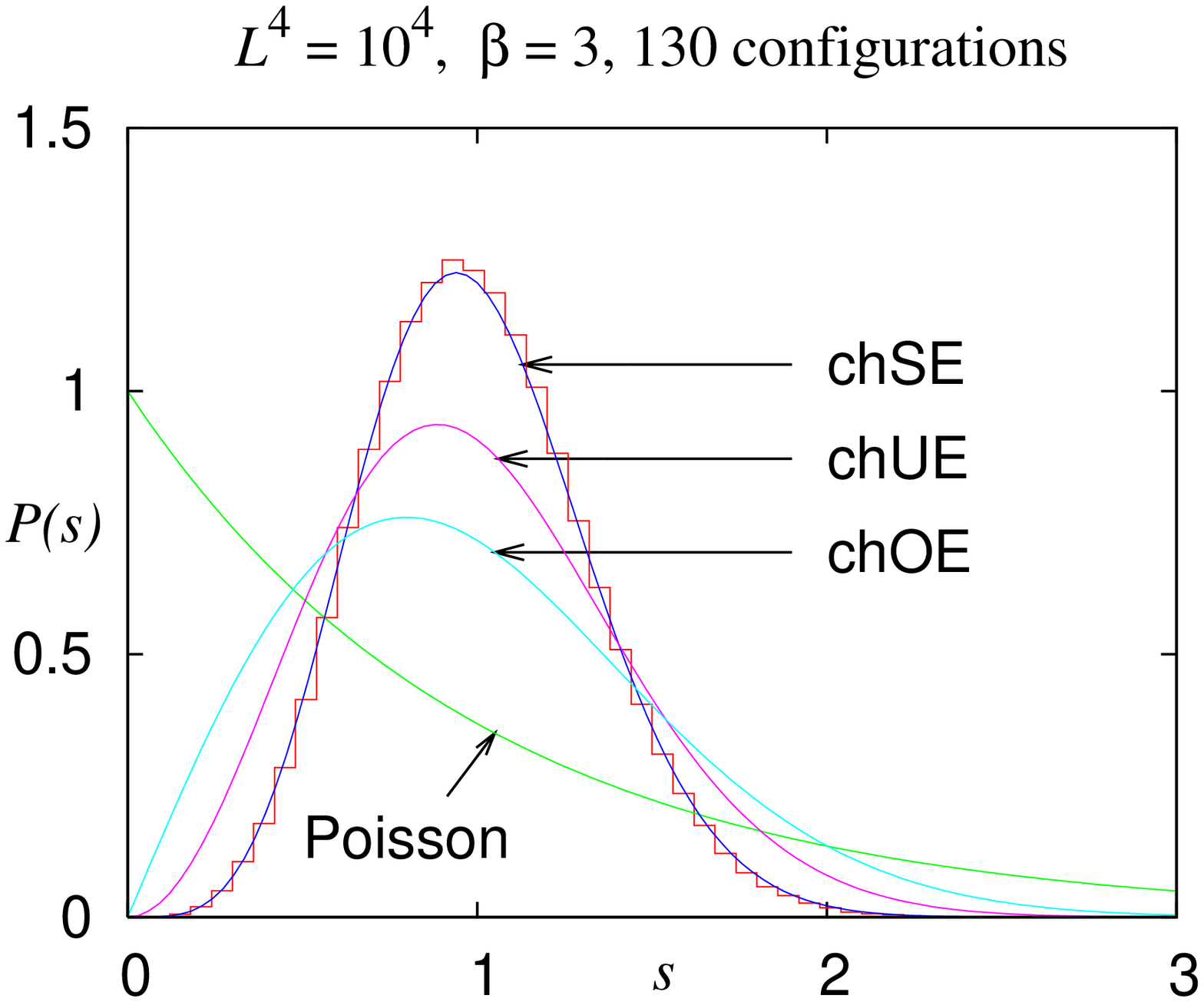}
  \includegraphics[width=.33\textwidth]{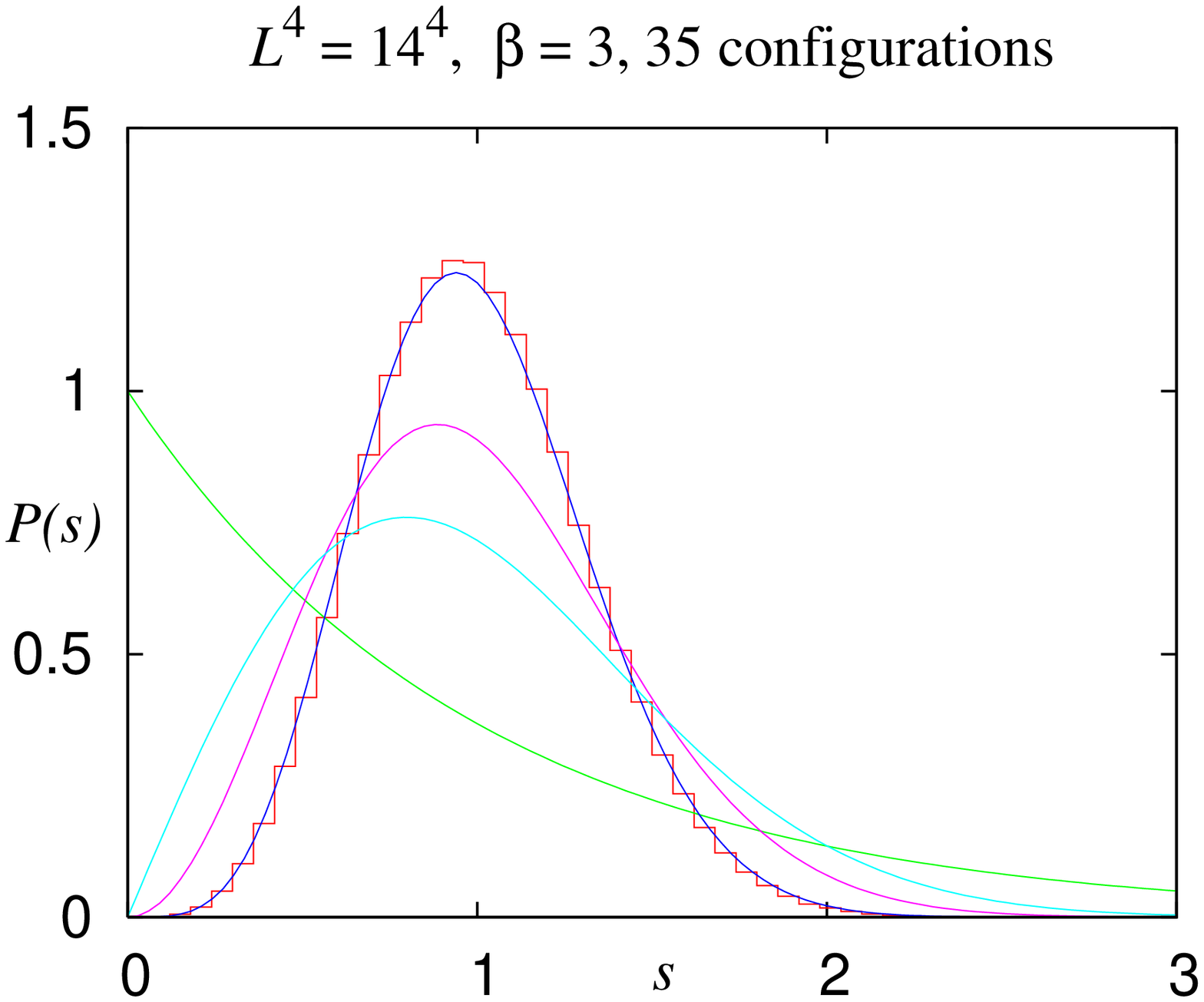}
  \includegraphics[width=.33\textwidth]{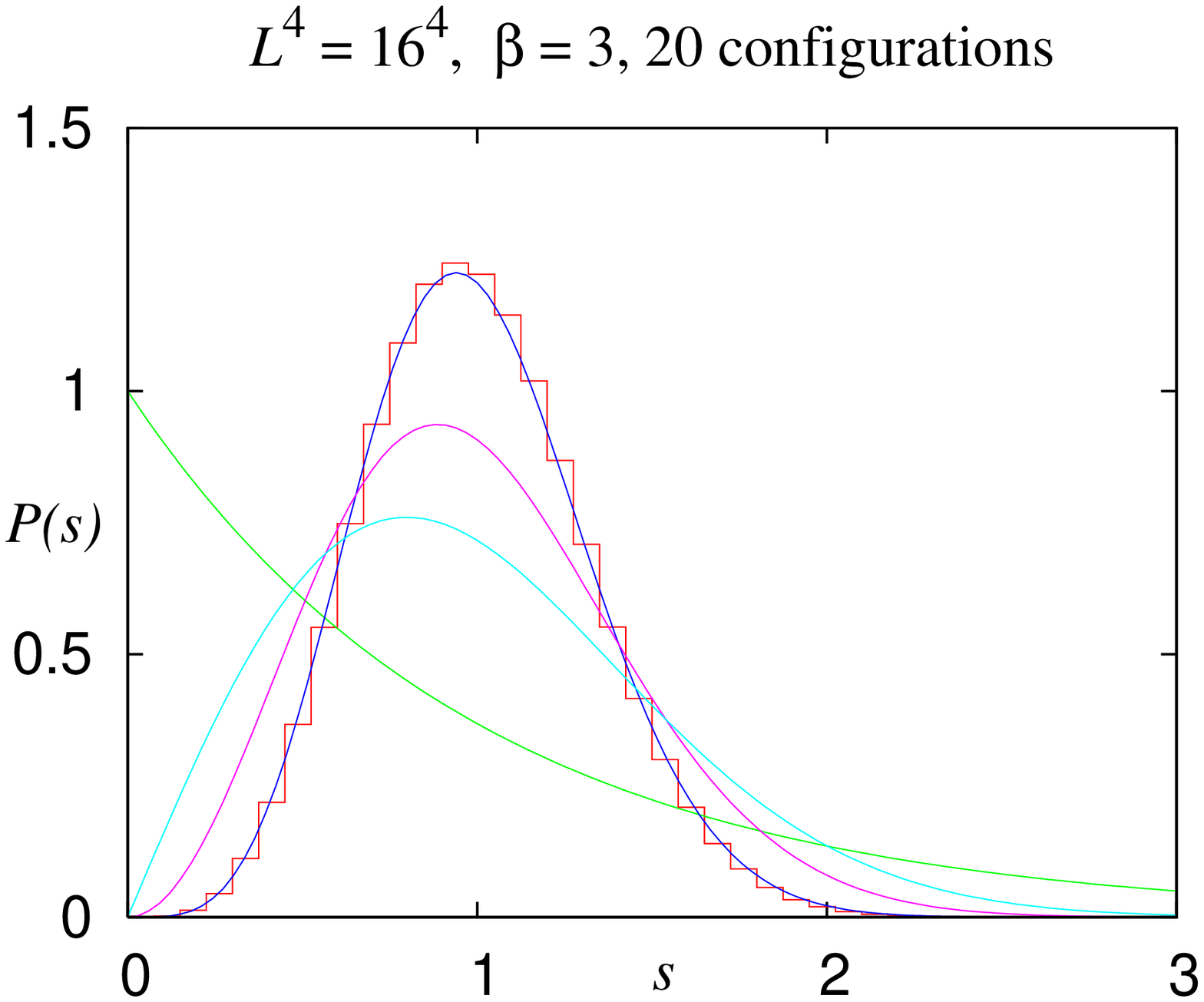}
  \caption{The distribution of the unfolded level spacing $s$ obtained
    for different lattice sizes from the SU(2) staggered Dirac
    operator with fundamental fermions and gauge action parameter
    $\beta=4/g^2=3.0$ (histograms) is consistent with the chSE, which
    is not the symmetry of the continuum Dirac operator.}
  \label{fundsmallbetafig}
\end{figure}

The massless staggered Dirac operator on a lattice in $d$ dimensions
with lattice spacing $a$,
\begin{equation}
  \label{eq:DKS}
  (D_\text{KS})_{x,y} = \frac{1}{2a} \sum_{\mu=1}^{d}
  (-1)^{\sum\limits_{\nu<\mu}\!\! x_\nu} \left[ \delta_{x+\hat\mu,y} 
    U_\mu^\dagger(x) - \delta_{x-\hat\mu,y} U_\mu(x-\hat\mu) \right] \: , 
\end{equation}
which is widely used in numerical simulations, exhibits a peculiar
feature: For gauge group SU(2) and fundamental fermions, its
antiunitary symmetry is that of the chSE \cite{Halasz:1995vd} instead
of the chOE symmetry of the continuum operator.\footnote{A similar
  situation occurs for adjoint fermions: The staggered Dirac operator
  has chOE symmetry in this case, as opposed to the chSE symmetry of
  the continuum operator.}  This discrepancy is due to the replacement
of the $\gamma$-matrices by the staggered phases in $D_\text{KS}$.
Fig.~\ref{fundsmallbetafig} confirms the expectation using $P(s)$ as
an example.

A transition of the symmetry properties of the staggered Dirac
operator from chSE to chOE is expected in the continuum limit.  A
first indication of such a transition has been reported in
Ref.~\cite{Follana:2006zz}.  We plan to study this transition in more
detail, but here we first consider a numerically cheaper case, namely
the free limit.  This limit is approached by increasing $\beta$ at
fixed (or mildly varying) lattice size, i.e., the physical volume is
shrinking to zero.  From the RMT point of view, this limit is
interesting since it might result in a transition to Poisson behavior
in, e.g., $P(s)$.  We shall see that the situation is actually a bit
more complicated.

\section{The Dirac spectrum of vacuum configurations and the influence
  of Polyakov loops}
\label{sec:vac}

In this section we present a short theoretical interlude in
preparation for our numerical results.  Let us consider a particular
gauge configuration.  For reasons that will become clear below, we now
construct a corresponding vacuum configuration (i.e., a configuration
with all plaquettes equal to unity) that is built from uniform links
in an Abelian subgroup of SU(2) which reproduce the average traced
Polyakov loops $P_\mu$ (for all directions $\mu$) of the configuration
under consideration.  The eigenvalue spectrum of the staggered Dirac
operator \eqref{eq:DKS} can be computed analytically for such a vacuum
configuration.  If the lattice extent in the $\mu$-direction is
$L_\mu$, we obtain
\begin{equation}
  \label{clustersandpolyakovloops}
  \lambda=\pm i \sqrt{\sum_{\mu=1}^d \sin^2 \left[ \frac{2 \pi}{L_\mu}
      \left( k_\mu + c_\mu + \frac{\arccos P_\mu}{2 \pi} \right)\right] }
  \quad \text{with }\: k_\mu\in\mathbb N\:,\; 0 \le k_\mu < \frac{L_\mu}2\:,
\end{equation}
where $c_\mu = 0$ $\left(c_\mu= \frac{1}{2}\right)$ for (anti-)
periodic boundary conditions (b.c.s) of the Dirac operator in
direction $\mu$.  In the free limit, i.e., for $\beta \to \infty$, the
Polyakov loops take values in the center $\mathbb Z_2$ of SU(2), i.e.,
$P_\mu=\pm1$.  In this case it is clear from
Eq.~\eqref{clustersandpolyakovloops} that changing $P_\mu$ from $+1$
to $-1$, or vice versa, is equivalent to switching between periodic
and antiperiodic b.c.s in that direction.  In the following we always
use (anti-) periodic b.c.s for $\mu=1$, 2, 3 ($\mu=4$).  Close to the
free limit, i.e., for large values of $\beta$, the distribution of
$P_\mu$ is peaked at $\pm1$.  The eigenvalues predicted by
Eq.~\eqref{clustersandpolyakovloops} are degenerate, see below.

\section{Numerical results for the eigenvalue spectrum close to the
  free limit}

\begin{figure}[-t]
  \includegraphics[width=.33\textwidth]{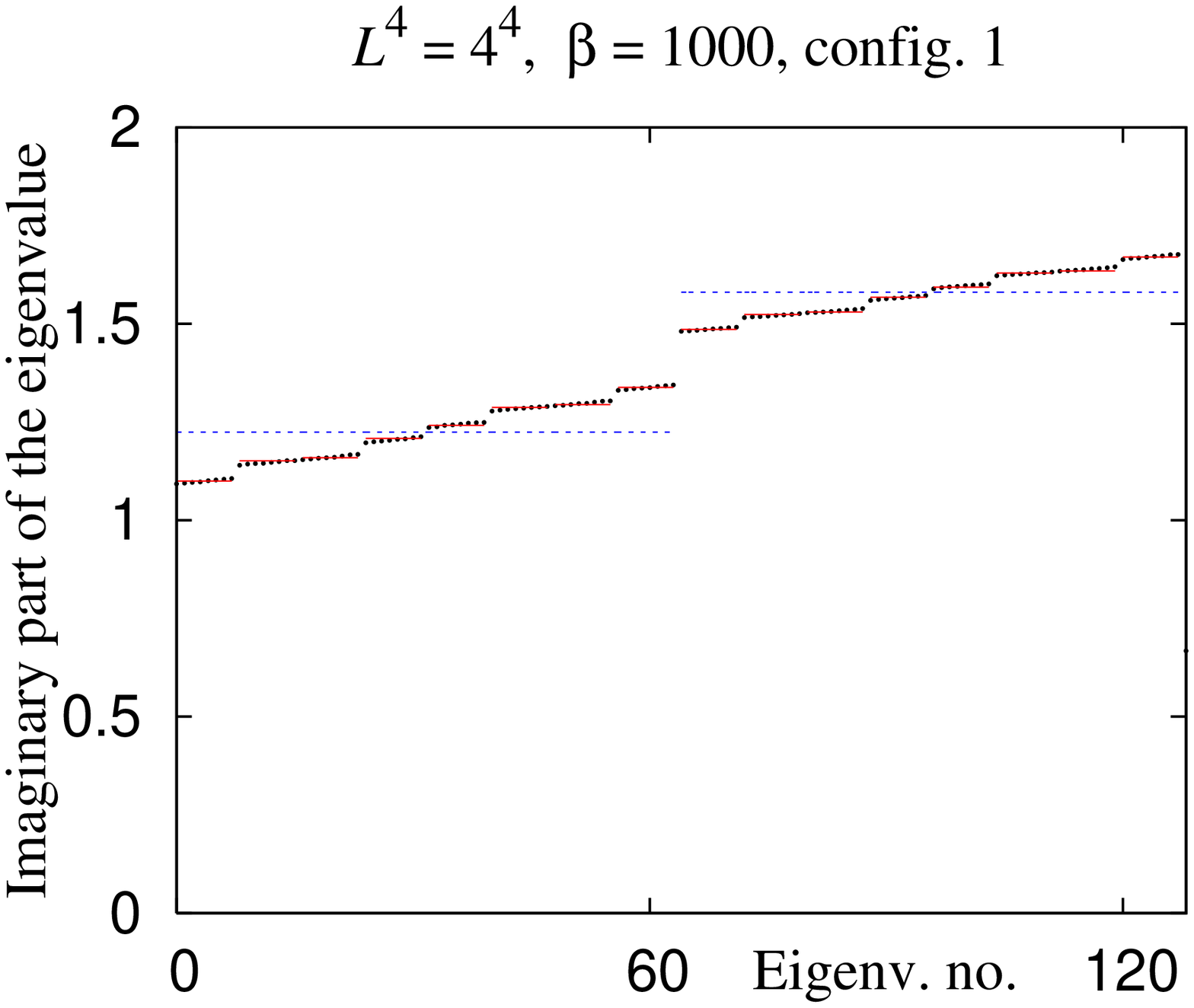}
  \includegraphics[width=.33\textwidth]{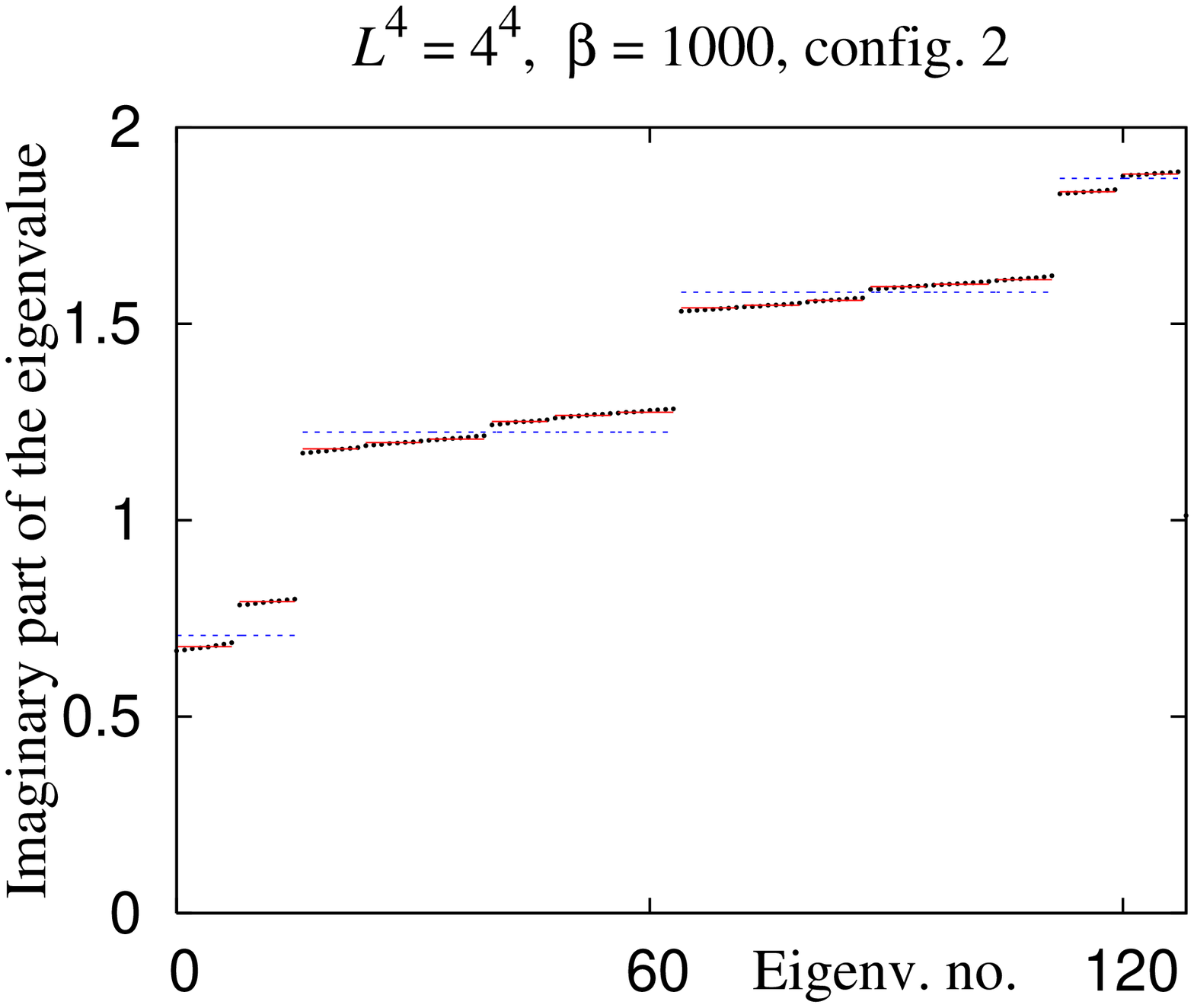}
  \includegraphics[width=.33\textwidth]{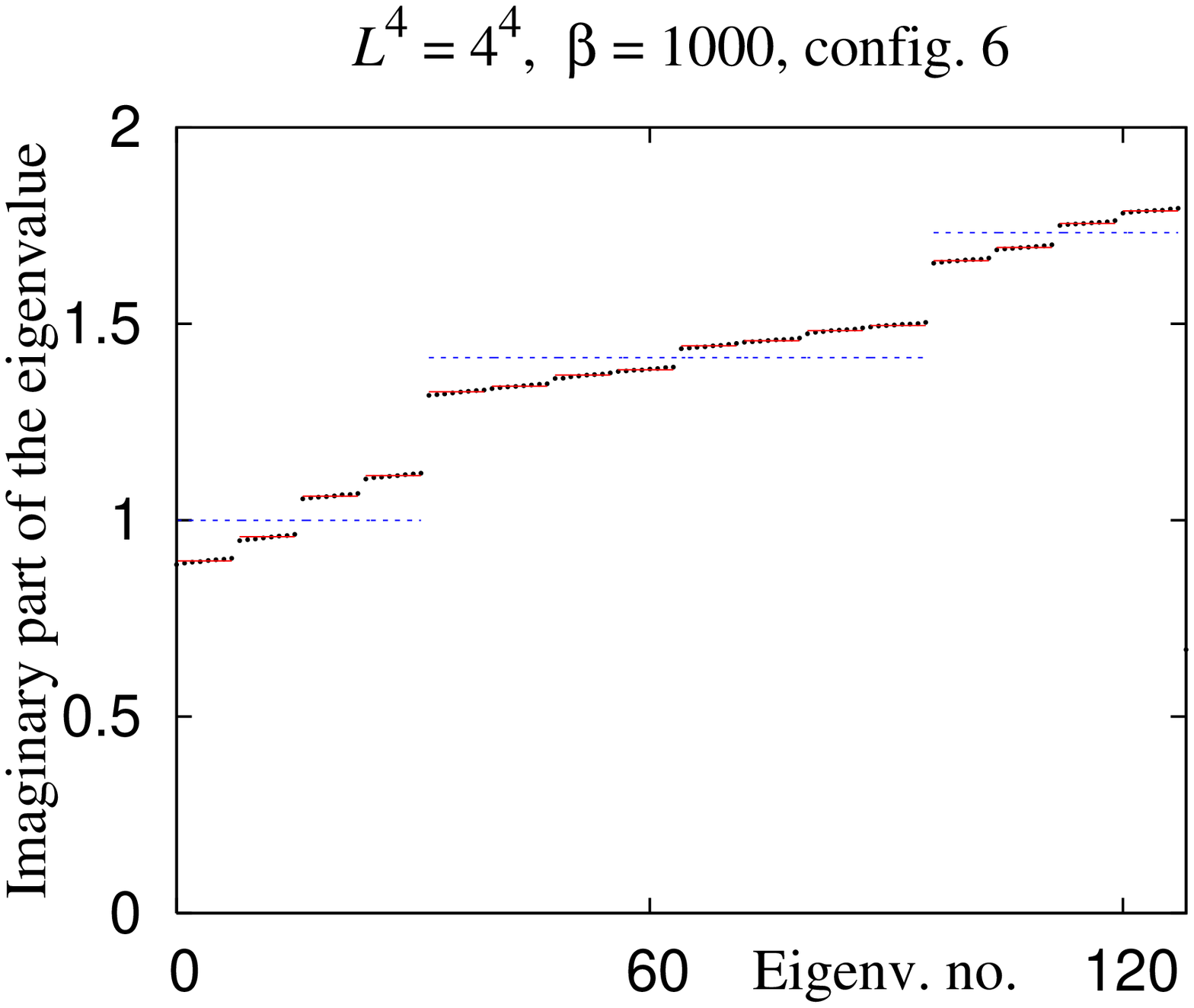}
  \includegraphics[width=.33\textwidth]{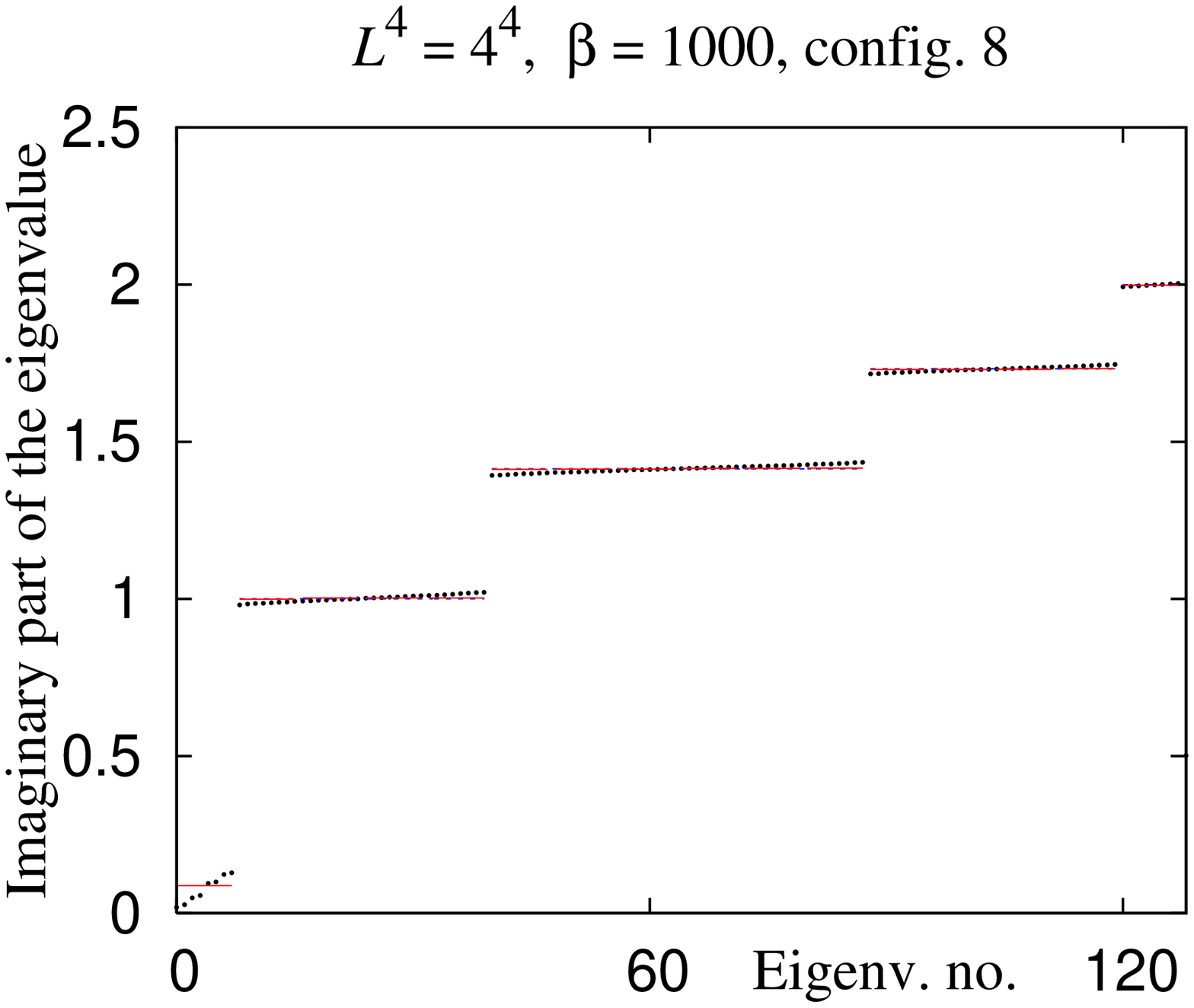}
  \includegraphics[width=.33\textwidth]{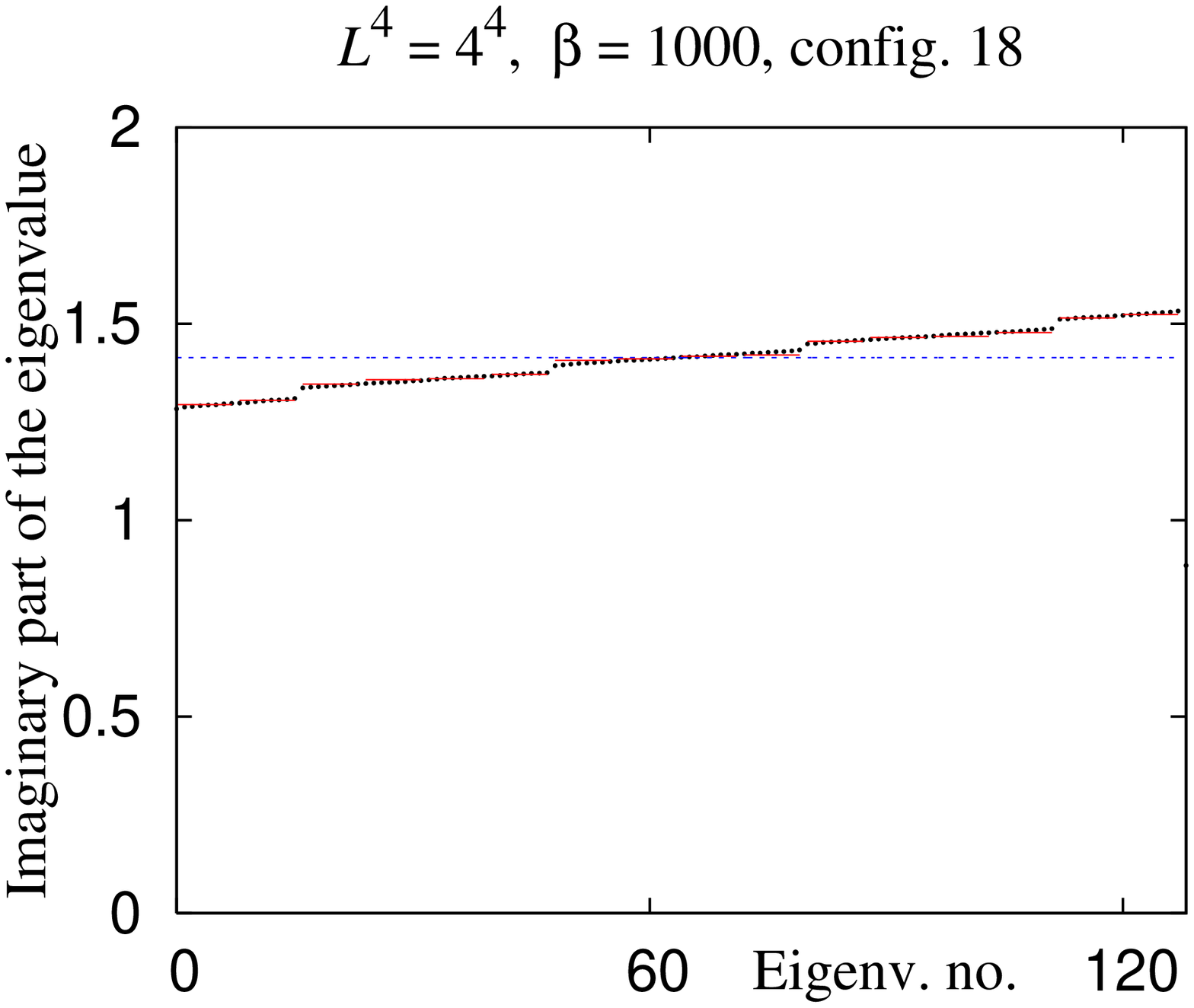}
  \includegraphics[width=.33\textwidth]{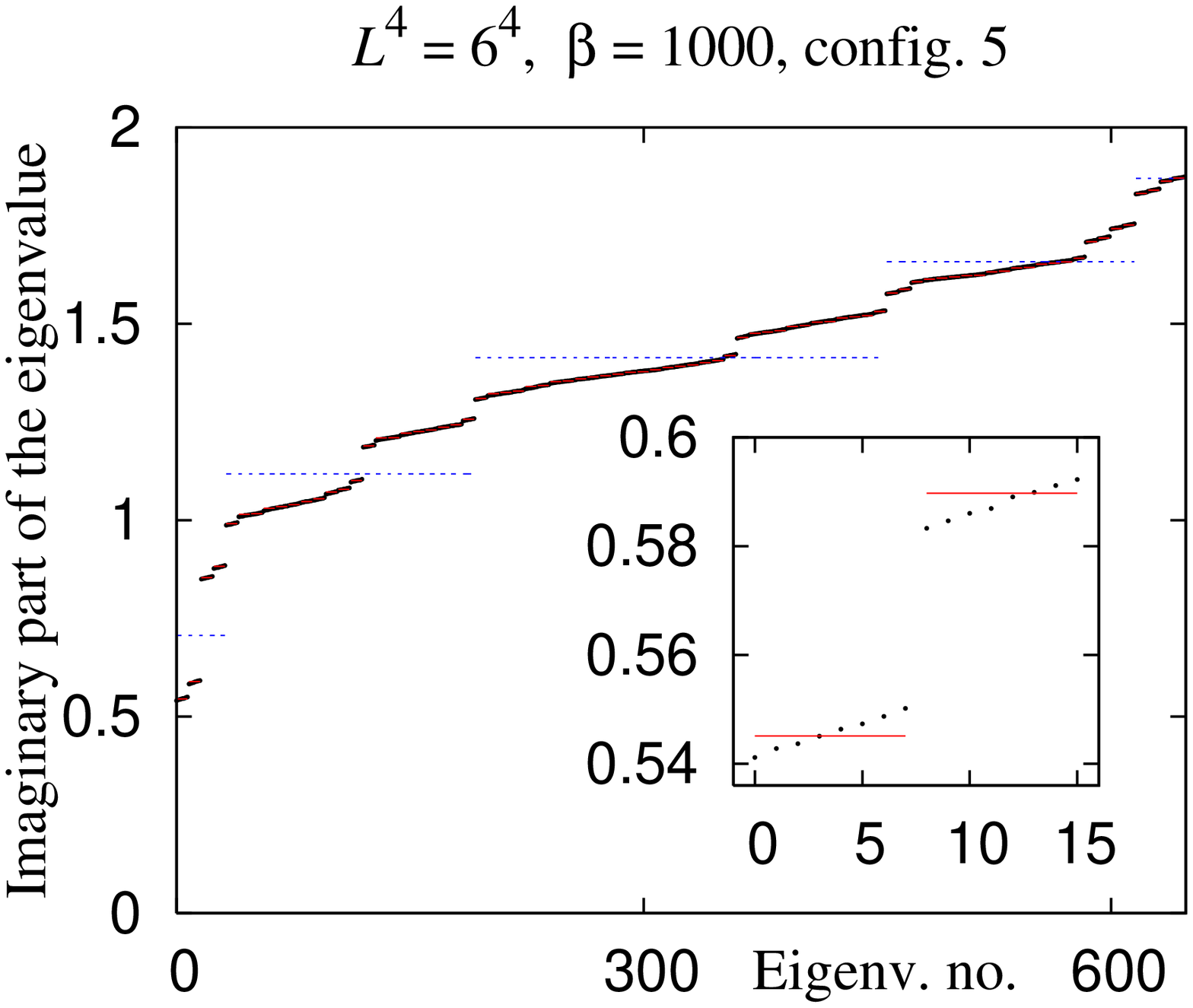}
  \caption{Separation of scales in the level spacings close to the
    free limit.  The eigenvalues obtained for each configuration
    (black dots) arrange themselves in clusters of eight.  These
    clusters are spread about the well-separated plateaux
    corresponding to the free case (dashed blue lines).
    Eq.~\protect\eqref{clustersandpolyakovloops} yields an accurate
    prediction for the location of each cluster (solid red
    lines). The different plateau structures are due to the different
    signs of $P_\mu\approx\pm1$ in each configuration.  The last plot
    confirms that the agreement between the data and
    Eq.~\protect\eqref{clustersandpolyakovloops} persists on larger
    lattices, for which the theoretical formula predicts more
    clusters.}
  \label{threescalesfig}
\end{figure}

When $\beta$ is increased to very large values, we observe that the
eigenvalue spectrum of $D_\text{KS}$ arranges itself as shown in
Fig.~\ref{threescalesfig}.  Only the eigenvalues with positive
imaginary part are plotted, and an overall double (Kramers) degeneracy
\cite{Hands:1990wc} has been divided out from all of our results.  The
dashed blue lines, which will be called \emph{plateaux} in the
following, correspond to the highly degenerate eigenvalues predicted
by Eq.~\eqref{clustersandpolyakovloops} in the free limit, i.e., for
$P_\mu=\pm1$.  The numerically obtained eigenvalues form
\emph{clusters} consisting of eight eigenvalues each, and these
clusters are located close to the plateaux of the free limit.  We
observe a clear separation of three energy scales (from largest to
smallest),
\begin{enumerate}\itemsep-1mm
\item the spacings between the plateaux of the free limit,
\item the spacings between adjacent clusters (which, by definition, do
  not overlap), and
\item the spacings between adjacent eigenvalues within a cluster.
\end{enumerate}

The question now arises to what extent the locations of the clusters
for a particular configuration can be described by the levels obtained
from Eq.~\eqref{clustersandpolyakovloops} for the vacuum configuration
constructed as described in Sec.~\ref{sec:vac}.  The answer is given
by the solid red lines, which correspond to the predictions of
Eq.~\eqref{clustersandpolyakovloops} using the $P_\mu$ computed from
the configuration under consideration.  These lines are essentially
hidden by the data points and thus give very good approximations to
the cluster locations.  This statement holds for all configurations,
some of which are shown in Fig.~\ref{threescalesfig}.

We can understand the observation that the clusters contain eight
nearly degenerate eigenvalues.  A careful analysis shows that the
eigenvalues predicted by Eq.~\eqref{clustersandpolyakovloops} have a
multiplicity of $2^d$. For $d=4$, this predicts an eightfold
degeneracy in addition to Kramers' degeneracy.  A small perturbation
of the vacuum configuration lifts this eightfold degeneracy but not
the Kramers degeneracy, which is exact for $D_\text{KS}$.

We also remark that in the continuum limit, in which the lattice
spacing goes to zero at fixed physical volume, the eigenvalues of
$D_\text{KS}$ should arrange themselves in multiplets corresponding to
the taste degeneracy of staggered fermions.  This effect was observed
for SU(3)~\cite{quadruplets} (quadruplets) and
SU(2)~\cite{Follana:2006zz} (doublets) for improved versions of
$D_\text{KS}$, but it should not be confused with the effect we are
studying here.

Our observations may be related to other recent work~\cite{chisbconf}
discussing the connection between the spectrum of the Dirac operator
(which is relevant for chiral symmetry breaking) and the Polya\-kov
loop (which is an order parameter for confinement in the quenched
theory).

\section{Spectral fluctuations on different scales}

We now turn to a study of spectral correlations close to the free
limit, using again the nearest-neighbor spacing distribution $P(s)$ as
an example.  To construct $P(s)$, the average spectral density must be
separated from the spectral fluctuations by an unfolding procedure.
Because of the separation of scales observed above, a uniform
unfolding of the entire spectral density is not sensible close to the
free limit.  Rather, we should consider the spectral fluctuations
separately on the three scales we identified.

\begin{figure}[t]
  \centering
  \includegraphics[width=.33\textwidth]{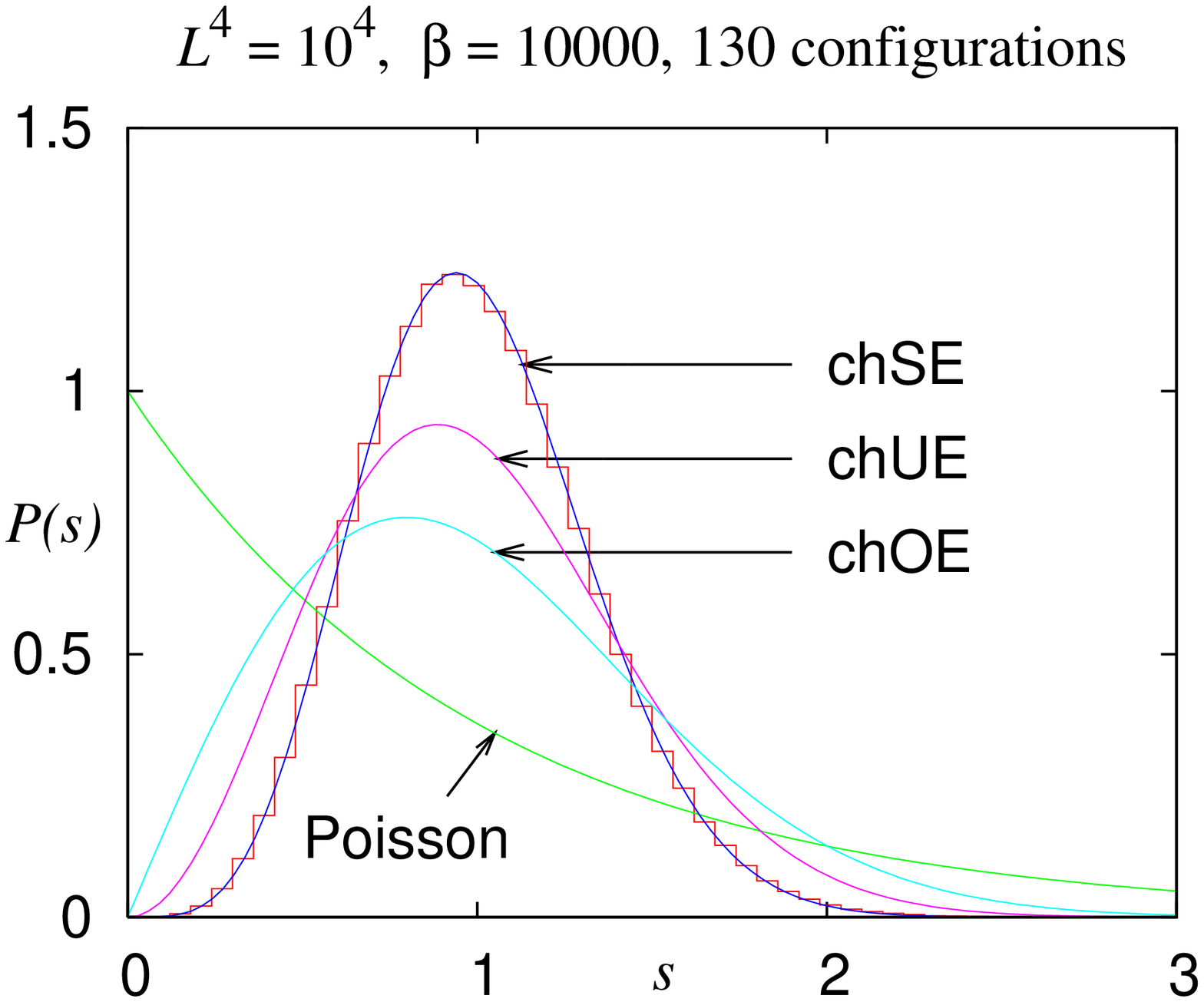}
  \hspace*{20mm}
  \includegraphics[width=.33\textwidth]{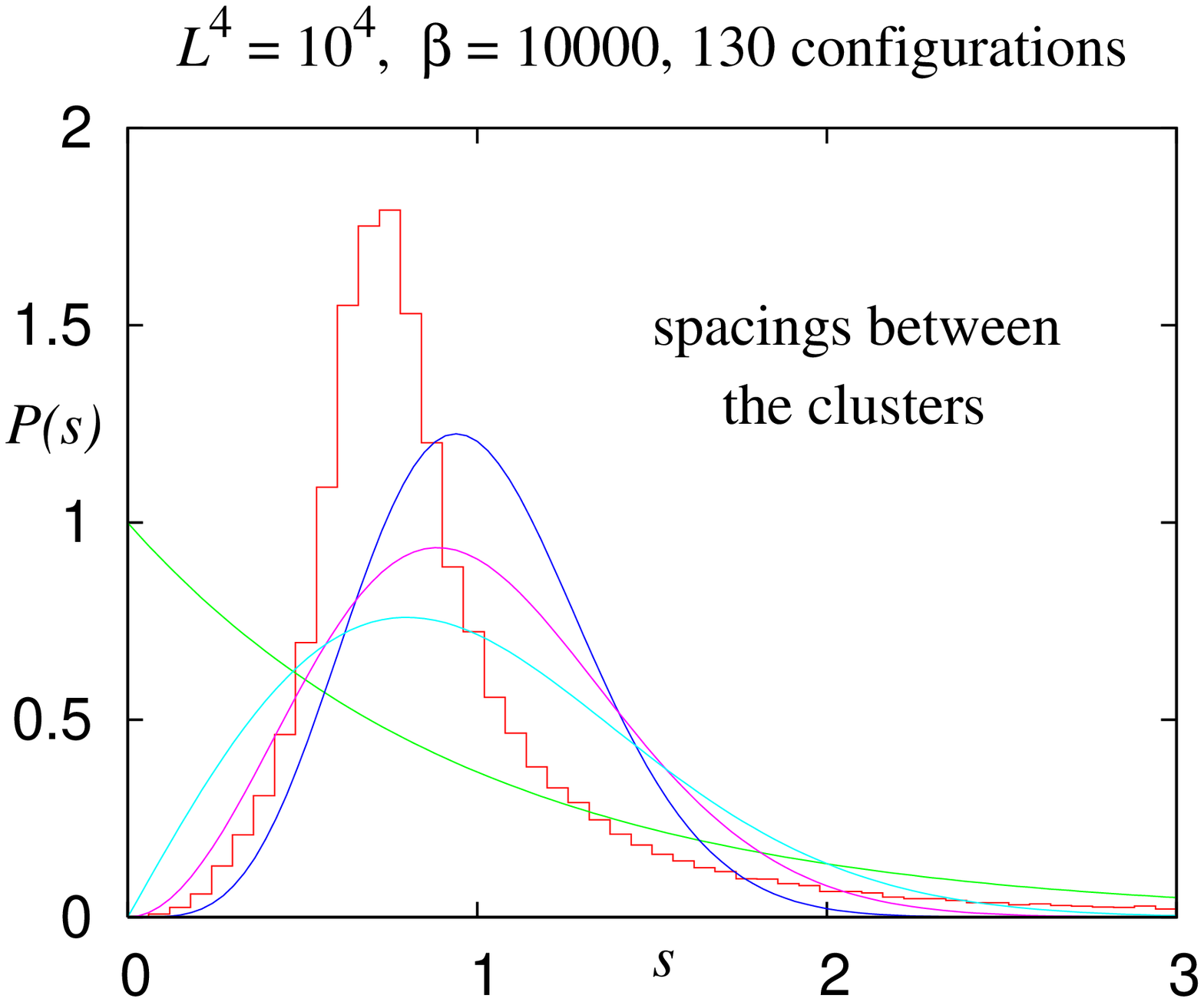}
  \caption{$P(s)$ for the eigenvalue spacings within the clusters
    (left) and for the spacings between clusters (right), both for
    $L^4=10^4$ and $\beta=10000$.}
  \label{still_chse}
\end{figure}

First, we construct $P(s)$ for the level spacings within the clusters
by unfolding the spectral density only within a given cluster and then
averaging $P(s)$ over all clusters.  Fig.~\ref{still_chse} (left)
shows that $P(s)$ within the clusters continues to agree with the chSE
even for very large values of $\beta$.  This is consistent with the
theoretical expectation, since the perturbation that lifts the
degeneracy of the eigenvalues in each cluster has the same symmetries
as the full $D_\text{KS}$ operator, which are those of the chSE.

\begin{figure}[t]
  \centering
  \includegraphics[width=.33\textwidth]{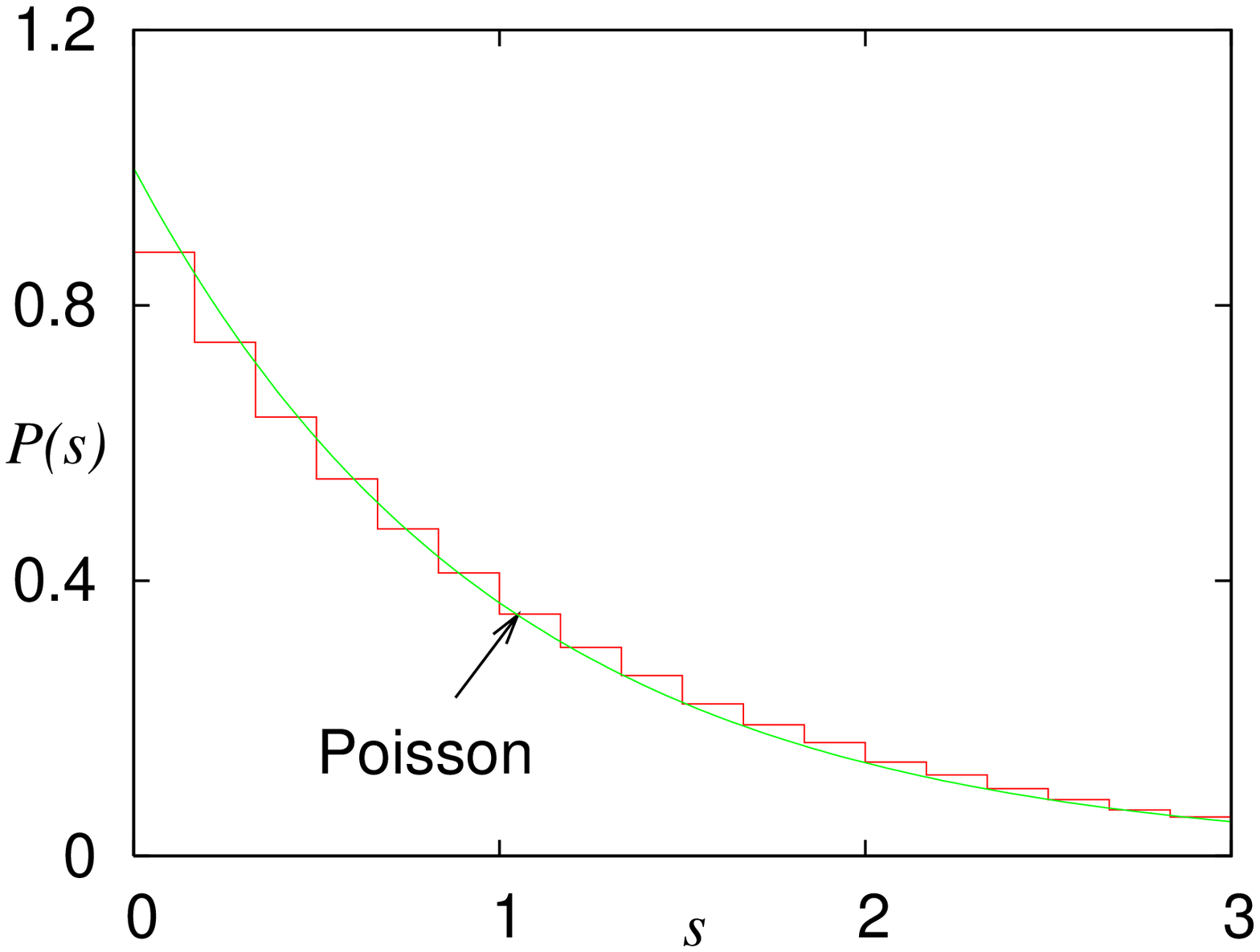}
  \hspace*{20mm}
  \includegraphics[width=.33\textwidth]{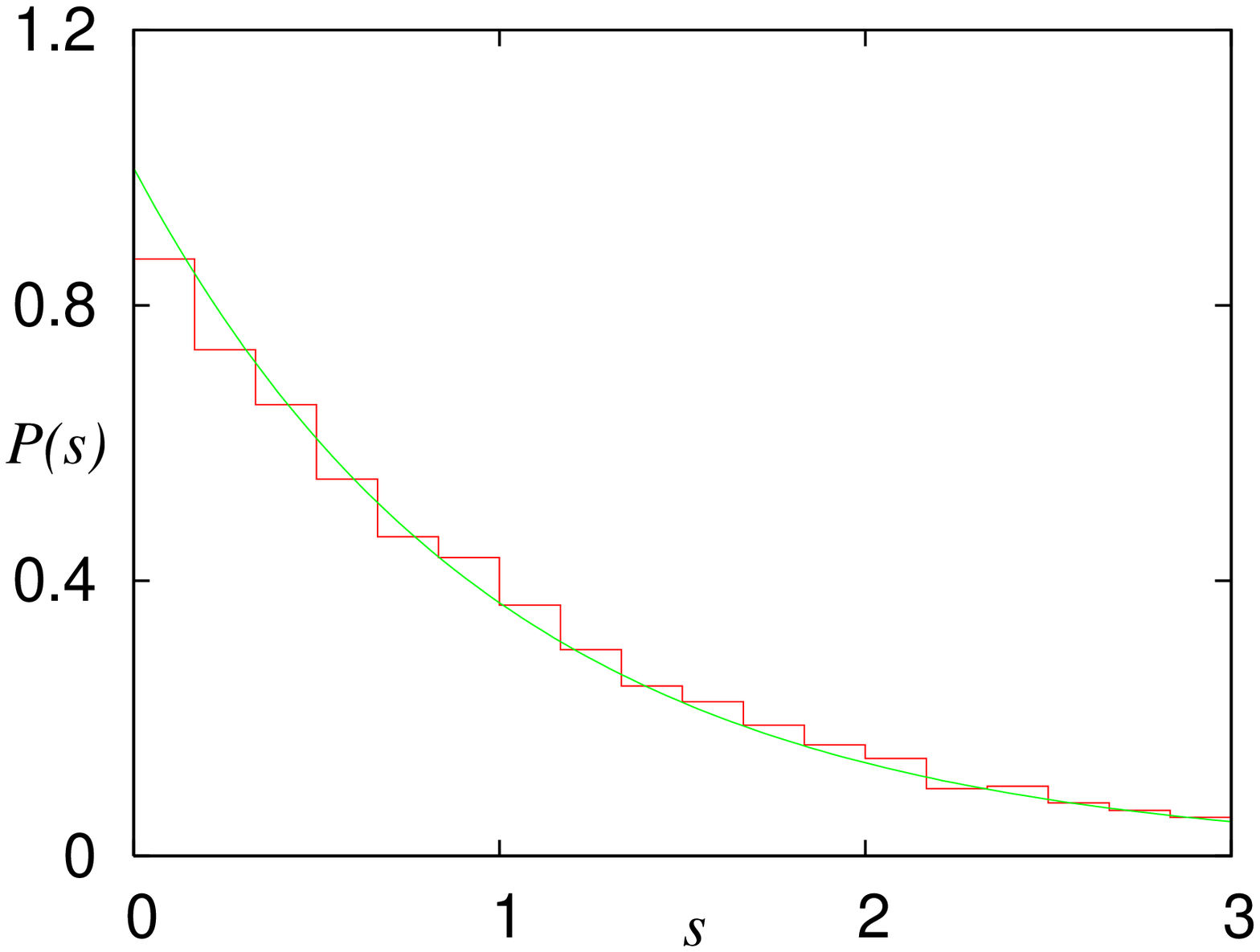}
  \caption{$P(s)$ for the ``cluster spectrum'' predicted by
    Eq.~\protect\eqref{clustersandpolyakovloops} for a single
    configuration on a $ 34 \times 38 \times 46 \times 58 $ lattice at
    $\beta=10000$ (left) and $P(s)$ for the free Dirac eigenvalues (or
    plateaux) on the same lattice (right).  Both agree with the
    Poisson distribution.}
  \label{poissonlargelatticefig}
\end{figure}

Second, to construct $P(s)$ for the spacings between clusters, we
define a cluster by the average of its eight members and unfold the
density of the clusters.  Fig.~\ref{still_chse} (right) shows that the
resulting $P(s)$ differs from the chSE.  It also differs from the
Poisson distribution, but we believe that this is due to the small
lattice size and to the fact that on a $10^4$ lattice the free
staggered operator has many ``accidental'' degeneracies.  These
degeneracies can be removed by choosing a lattice with
$L_\mu=2\ell_\mu$, where the $\ell_\mu$ are four different prime
numbers.  As an example, we generated quenched configurations close to
the free limit ($\beta=10000$) for a $34\times38\times46\times58$
lattice, computed the averaged traced Polyakov loops $P_\mu$ and used
these to calculate the ``cluster spectrum'' according to
Eq.~\eqref{clustersandpolyakovloops}.  The resulting $P(s)$ is shown
in Fig.~\ref{poissonlargelatticefig} (left) and now agrees with the
Poisson distribution.

Third, we consider $P(s)$ for the spacings between the free
eigenvalues (or plateaux), which are known analytically
(Eq.~\eqref{clustersandpolyakovloops} with $P_\mu=\pm1$).  Again it is
sensible to remove accidental degeneracies by choosing a ``prime
lattice''.  The result for a $34\times38\times46\times58$ lattice,
obtained after unfolding the free eigenvalues, is shown in
Fig.~\ref{poissonlargelatticefig} (right) and agrees with Poisson as
expected \cite{GoesToPoisson}.

Although the two plots in Fig.~\ref{poissonlargelatticefig} look very
similar, it should be noted that they come from data at very different
scales.  The average spacing between the levels of the free spectrum
is more than ten times larger than the average spacing between the
levels predicted by Eq.~(\ref{clustersandpolyakovloops}).

\section{Summary and outlook}

We have investigated the spectrum of the staggered Dirac operator with
SU(2) gauge fields close to the free limit.  Three different energy
scales emerge:
\begin{enumerate}
\item Overall plateau structure: The spectrum arranges itself in
  clusters of eight eigenvalues each, lying close to the plateaux
  predicted for the free Dirac operator. The plateau structure only
  depends on the lattice geometry (i.e., on the $L_\mu$ and on the
  b.c.s) and on the signs of the average traced Polyakov loops in the
  different directions.  (Note that the distribution of the traced
  Polyakov loops is peaked at $\pm1$, corresponding to the center
  elements of SU(2).)
\item Plateau-breaking and cluster separation at an intermediate
  scale: At a finer scale, the spread of the clusters about the
  plateaux of the free limit is due to the deviations of the $P_\mu$
  from $\pm1$ and can be accurately modeled by
  Eq.~\eqref{clustersandpolyakovloops}.
\item Eigenvalue splitting within the clusters: The system dynamics
  removes the degeneracy of the eight eigenvalues belonging to the
  same cluster.
\end{enumerate}
In the regime we have studied, these three scales are well separated
and can be unambiguously disentangled from each other.

The nearest-neighbor spacing distribution $P(s)$ computed within the
clusters shows a behavior compatible with the chSE, consistent with
the symmetries of the staggered Dirac operator.  For large enough
``prime lattices'', the spacing distributions between the clusters and
between the plateaux tends to the Poisson distribution.  In the near
future, we will also present a study of the spectrum of the Dirac
operator for adjoint fermions close to the free limit.  Ultimately, of
course, we would like to obtain a more detailed understanding of the
continuum limit, in which a chSE to chOE (for SU(2) with fundamental
fermions) or chOE to chSE (for adjoint fermions) transition is
expected.

\acknowledgments

We thank J.J.M.~Verbaarschot for helpful discussions and acknowledge
support from DFG (FB, SK, TW) and from the Alexander von Humboldt
Foundation (MP).

\end{document}